\documentclass[a4paper]{llncs}

\usepackage{amssymb}
\usepackage{graphicx}
\usepackage{xspace}
\usepackage{color}
\usepackage{mdwlist}
\usepackage{listings}
\usepackage{subcaption}
\usepackage[htt]{hyphenat}
\usepackage{hyperref}
\usepackage[footnote,nomargin,noinline,nomarginclue,draft]{fixme}

\FXRegisterAuthor{vvdv}{avvdv}{\textcolor{red}{VVDV}}

\FXRegisterAuthor{hjb}{ahjb}{\textcolor{red}{HJB}}

\FXRegisterAuthor{vm}{avm}{\textcolor{blue}{Vee}}

\usepackage{url}

\urldef{\mailsa}\path|r.k.konoth@vu.nl, herbertb@cs.vu.nl|
\urldef{\mailsaa}\path|R.S.vanWegberg@tudelft.nl|
\urldef{\mailsaaa}\path|email@veelasha.org|

\newcommand{\keywords}[1]{\par\addvspace\baselineskip
\noindent\keywordname\enspace\ignorespaces#1}

\let\OldTexttt\texttt
\renewcommand{\texttt}[1]{{\fontseries{b}\selectfont \OldTexttt{#1}}}

\def\Cc{Cryptocurrency}

\def\cc{cryptocurrency}

\def\ccs{cryptocurrencies}

\renewcommand{\tabcolsep}{0.5em}

\begin{document}

\mainmatter  


\title{Malicious cryptocurrency miners: \\Status and Outlook}


%
%
\author{Radhesh Krishnan Konoth\inst{1} \and  Rolf van Wegberg\inst{2} \and \\
Veelasha Moonsamy\inst{3}  \and Herbert Bos\inst{1}\\}

\institute{Vrije Universiteit Amsterdam, The Netherlands \\
\mailsa\\
\and Delft University of Technology, The Netherlands \\
\mailsaa\\
\and Radboud University, The Netherlands \\
\mailsaaa\\}
%


%
%

\maketitle

\definecolor{dkgreen}{rgb}{0,0.6,0}
\definecolor{gray}{rgb}{0.5,0.5,0.5}
\definecolor{mauve}{rgb}{0.58,0,0.82}

\begin{abstract}
In this study, we examine the behavior and profitability of modern
malware that mines cryptocurrency. Unlike previous studies, we look at
the cryptocurrency market as a whole, rather than just Bitcoin. We not only 
consider PCs, but also mobile phones, and IoT devices. In the past
few years, criminals have attacked all these platforms for the purpose
of cryptocurrency mining. The question is: how much money do they make? It is
common knowledge that mining Bitcoin is now very difficult, so why do
the criminals even target low-end devices for mining purposes? By
analyzing the most important families of malicious cryptocurrency miners that were active
between 2014 and 2017, we are able to report how they work, which
currency they mine, and how profitable it is to do so. We will see
that the evolution of the cryptocurrency market with many new
cryptocurrencies that are still CPU minable \emph{and} offer better
privacy to criminals and have contributed to making mining malware
attractive again---with attackers generating a continuous stream of
profit that in some cases may reach in the millions.

\keywords{malware, cryptocurrency mining, mobile phone, IoT}
\end{abstract}

\section{Introduction}


A cryptocurrency is a digital asset designed to work as a medium of
exchange.  Bitcoin~\cite{nakamoto} became the first decentralized
cryptocurrency in 2009. By design, \ccs{} need significant
computational processing to validate transactions and add them to a
distributed ledger (the \emph{blockchain}), and networks of so-called
miners therefore set themselves to the task of maintaining the working
of Bitcoin. Incentivized by financial reward, cryptocurrency miners
uphold the network by validating transactions. The financial reward
serves as a compensation for the computing power needed to execute the
aforementioned tasks. However, if one could steal or borrow computing
power from others, the financial reward would grow significantly,
since the system would generate income at little or no cost to the
benificiary. This is what motivated cyber criminals to experiment, as
early as 2009, with using botnets of infected machines as
\emph{silent} cryptocurrency miners to maximize their profits.

For some time, in the days that banking trojans such as GameOver Zeus
ruled cyber crime~\cite{zeus}, cryptocurrency mining gained a certain
amount of popularity among cyber criminals who were, after all,
already in the business of compromising PCs and herding large numbers
of them in botnets.  In those days, we saw the first criminals
infecting machines exclusively to steal CPU resources to mine Bitcoins
on their behalf. However, after a few years already, the profits of
the Bitcoin mining botnets dwindled as the mining became too difficult
for regular machines and Bitcoin mining botnets fell into decline.
Analyses by security companies in 2014 suggested that malicious miners
are not profitable on PC and certainly not on mobile
devices~\cite{mcafee_threat_report:2014}.


Banking trojans and later also ransomware became the cyber criminals'
workhorses. Both have drawbacks. For instance, cashout is tricky and
researchers have shown that it is possible to trace even Bitcoin
transactions~\cite{bitiodine,philip,sarah}. Switching to more
privacy-preserving currencies such as Monero or Zcash is not easy,
because in many countries it is not possible to buy such currencies
with normal bank accounts. Moreover, both banking trojans and
ransomware tend to be noisy---users notice when money is
stolen. After the theft, most bots are burned, as users clean up their
systems to prevent further damage.

Botnets of cryptocurrency miners have no such disadvantages. A bot can
easily mine privacy-preserving currencies, to make all profits
untraceable. The theft is also stealthy, as no money transfers from
the user's bank account. The cost to the user is in a higher
electricity bill and reduced computing performance.  Moreover, rather
than a one-time hit, the cryptocurrency miner can generate a
continuous stream of income.

Finally, one main cause of the decline of cryptocurrency malware has
disappeared: it is now profitable to mine cryptocurrencies with
regular CPUs again. New cryptocurrencies are introduced all the time
and, unlike Bitcoin, these are still mineable without specialized
hardware. In fact, in this paper, we will show that even low-end IoT
devices are interesting targets for crypto mining. The market now
counts over 1,500 cryptocurrencies, out of which more than 600 see
active trade. At the time of writing, they represent over 50\% of the
cryptocurrency market.

As many of these coins are CPU mineable and provide better privacy,
the research question we ask is the following: has 
cryptocurrency mining become attractive to cyber criminals again?

To answer this question, we perform the first scientific study of the
phenomenon of malicious cryptocurrency mining that goes beyond
Bitcoin. Like the McAfee report~\cite{mcafee_threat_report:2014} from
2014, a more narrow study by Huang et
al.~\cite{botcoin:ndss2014} of malicious Bitcoin miners between 2011 and
2013 found that Bitcoin mining was only marginally profitable. In the 3 years that followed, Bitcoin
mining has become so hard that without specialized hardware, it is no
longer practical at all. However, the world has changed and the number of alternative currencies has exploded. In a single sentence in the conclusions, Huang et al.~\cite{botcoin:ndss2014} speculated that such a change may make the mining
activities profitable again, but leave this for future work. In
this paper, we examine such alternative coins.  We will look at
malware in the wild by analyzing existing malicious cryptocurrency
miners targeting different platforms for different currencies, and
provide methods to detect them. We will see that in 2017 alone, four
new families of malicious cryptocurrency miners have emerged.  None of them mine
Bitcoin. Moreover, we investigate to what extent the evolution of the
cryptocurrency market made mining malware more practical and
profitable.  Our study shows that it makes economic sense for
attackers to invest in this type of activity.

\textbf{Contributions} We make the following contributions:

\begin{itemize}
\item  We study how the growth of the cryptocurrency market and the growth in computation power of devices favors malicious cryptocurrency miners.
\item  We asses the profitability of a criminal business models using malicious cryptocurrency miners, its likeliness of occurrence and impact.
\item  We analyze existing malicious cryptocurrency miners to understand how they spread, what payload they use and how the cashout happens.
\end{itemize}
  
\textbf{Outline} The rest of the paper is organized as follows. In
Section~\ref{sec:background}, we introduce and describe the basic
terminology of cryptocurrrency mining. Section~\ref{sec:dataset}
includes a detailed analysis of our data set of malicious
cryptocurrencies that exist in the wild.  A discussion on our findings
and, importantly, an outlook for the future and detection mechanisms are presented in
Section~\ref{sec:outlook}. In Section~\ref{sec:relatedwork}, we
provide a brief overview of the literature review and finally conclude
the paper in Section~\ref{sec:conclusion}.

 
%
%
%
%

\section{Background on Cryptocurrencies}
\label{sec:background}

A cryptocurrency is a medium of exchange much like the euro or the dollar,  except that
it uses cryptography and blockchain
technology to control the creation of monetary units and to verify the 
transaction of a fund. \emph{Bitcoin}~\cite{nakamoto} was the first such 
decentralized digital currency.  A
cryptocurrency
user can transfer money to another user by forming a transaction record 
and committing it to a distributed write-only database called 
\emph{blockchain}.  The \emph{blockchain} is maintained by a peer-to-peer network of \emph{miners}. 
A \emph{miner} collects transaction data from the network, validates it and inserts into the 
blockchain in the form of a block.  When a miner successfully 
adds a valid block to the blockchain, the network compensates the miner with
cryptocurrency (e.g., Bitcoins). In the case of Bitcoin, this process is 
called \emph{Bitcoin mining} and this is how new Bitcoins enter circulation. 
Bitcoin transactions are protected with cryptographic techniques that 
ensure only the rightful owner of a Bitcoin address can transfer funds from it.

Since Bitcoin was created in 2009,  around 1500 other types of cryptocurrencies have been 
introduced~\cite{ElBahrawyAKPB17}. We commonly refer to these cryptocurrencies  
as \emph{alternative coins}---\emph{altcoins}, for short. 
Like Bitcoin, altcoins also use the blockchain technology and have a similar reward mechanism. 
However, each altcoin differs in other characteristics, such as speed, 
traceability, and security. For instance, the Monero altcoin provides more privacy 
than any other currently existing cryptocurrency. With major industrial 
players such as J.P Morgan Chase, Microsoft, Intel, and Google backing some of 
these cryptocurrencies, altcoins are increasingly popular.
Today, altcoins such as  Ethereum, Ripple, Litecoin, Dash and Monero together make up a little over 
50 percent of the total cryptocurrency market~\cite{coinmarketcap}
while the rest is owned by Bitcoin. 

\subsection{\Cc{} Mining}

To add a block (i.e., a collection of transaction data) to the
blockchain, a miner has to solve a cryptographic puzzle based on the
block. This mechanism prevents malicious nodes from trying to add
bogus blocks to the blockchain and earn the reward illegitimately. A
valid block in the blockchain contains a solution to a cryptographic
puzzle that involves the hash of the previous block, the hash of the
transactions in the current block, and a wallet address to credit with
the reward.

In accordance with Satoshi Nakamoto's original Bitcoin
paper~\cite{nakamoto}, the puzzle is designed such that the
probability of finding a solution by a miner is proportional to the
computational power. Additionally, the difficulty of solving the
puzzle increases with the length of the blockchain.
Consequently, a situation arose where mining for Bitcoin using a
regular CPU was no longer profitable. Instead, miners started using,
specialized mining hardware in ASICs and FPGAs.

\subsection{\Cc{} Mining Pools}

As mentioned, the probability of mining a block is proportional 
to the computational resources used for solving the 
associated cryptographic puzzle. Due to the nature of the 
mining process, the interval between mining events exhibits 
high variance from the point of view of a single miner. 
In other words, a single home miner using a dedicated ASIC is unlikely to 
mine a block for years. Consequently, miners typically organize 
themselves into mining pools. All members of a pool work together 
to mine each block, and share the revenue when one of them successfully mines a block. 

The mining pool server assigns jobs to its members. 
To prove that a miner is contributing to solving the ultimate cryptographic puzzle,  
a miner submits this solution in the form of shares to the pool server.  
The pool server rewards the miner in proportion to the submitted number of valid shares.

\subsection{Pool Mining Protocol}
The protocol used by miners to 
reliably and efficiently fetch jobs from mining pool servers is known as Stratum~\cite{stratum}. It is a 
clear-text communication protocol built over TCP/IP, using a JSON-RPC 
format. Stratum prescribes that miners who want to join the mining pool first send a \emph{subscription} message, describing the miner's capability in terms of computational resources. 
The pool server then responds with a \emph{subscription response message}, 
and the miner sends an \emph{authorization request} message with its 
username and password. After successful authorization, 
the pool sends a \emph{difficulty notification} that is proportional to 
the capability of the miner---ensuring that low-end machines get easier jobs (puzzles) than high-end ones. Throughout this paper we will use the term \emph{high-end machine} to describe a PC and 
\emph{low-end machine} to describe both mobile devices and IoT devices, and Stratum ensures that even low-end machines may contribute to the mining process. 
Finally, the pool server assigns these jobs by means of  \emph{job notifications}. 
Once the miner finds a solution it sends the solution in the form of 
a \emph{share} to the pool server. The pool server rewards the miner in 
proportional to the number of valid shares it submitted and the difficulty of the jobs.


For instance, a user with a low-end machine will receive a low
difficulty, which means its  miner solves puzzles with low difficulty. A
high-end machine, on the other hand, should not solve such easy puzzles, because doing so
would overwhelm the pool server with large numbers of shares per
second. Instead, it will receive more difficult challenges that take
approximately the same time as easy challenges on the low-end machine. Irrespective of a machine's computational power, 
the miner's reward  will be proportional to the number of valid shares and their difficulty.

\section{Malicious Miners}
\label{sec:dataset}

We created a collection of malicious cryptocurrency miners for
different platforms (PC, mobile/Android, and IoT). Specifically, we
collected 197 samples of 8 different families, which we believe are
practically all the active families of malicious cryptocurrency miners
in our evaluation period. We received the samples from
VirusTotal\footnote{https://www.virustotal.com} (VT) and made a manual
effort to ensure that we did not miss any relevant families, by
analyzing as many blogs and forums that discuss crypto miners as we
could find~\cite{proofpoint,sophos} and downloading all the samples
based on the hashes mentioned in them. The majority of our dataset comprises of cryptocurrency miners that were active during the period of 2014--(August) 2017.

For PC platforms, we analyzed \texttt{BitcoinMiner.J}, \texttt{Mal/Miner-C}, \texttt{BitCoinMiner.hxao} and \texttt{Adylkuzz} that target Windows, as well as
\texttt{SambaCry} which targets Linux. For the mobile platform, we studied  samples of  the \texttt{Kagecoin} and \texttt{BadLepricon}  families, which are affecting Android users. Lastly, we added an IoT-based malicious cryptocurrency miner named \texttt{Shell.Miner}, which appeared in 2017 and affected a variety of IoT devices. Table~\ref{table:dataset} provides an overview of our dataset. 

\begin{table}[htp]
\begin{small}
  \caption{Our experimental dataset}
 \label{table:dataset}
  \centering
  \begin{tabular}{|p{2.7cm}|p{2.7cm}|c|c|}
    \hline
    \textbf{Family}&
    \textbf{Target Platform} & \textbf{VT: first seen} & \textbf{VT: last analyzed} \\
    \hline
    BitcoinMiner.J &
    PC (Windows) & 2009 & 2017 \\
    Mal/Miner-c &
    PC (Windows) & 2014 & 2017 \\
    Kagecoin &
    Android & 2014 & 2017 \\
    BitcoinMiner.hxao &   PC (Windows) & 2016 & 2017 \\
    BadLepricon &
    Android & 2017 & 2017 \\
    Adylkuzz &
    PC (Windows) & 2017 & 2017 \\
    Sambacry &
    PC (Linux) & 2017 & 2017 \\
    Shell.Miner &
    IoT & 2017 & 2017 \\
\hline
  \end{tabular}
\end{small}
\end{table}



\label{sec:Analysis}

In the remainder of this section, we  analyze the characteristics of the malware samples in our experimental dataset and discuss how these characteristics are evolving along with the cryptocurrency economy, as well as with the computation
power of both high- and low-end devices.

\subsection{Cryptocurrency mining on  PCs}
\texttt{BitcoinMiner.J}, \texttt{Mal/Miner-c},  \texttt{BitCoinMiner.hxao}, 
\texttt{Adylkuzz} and \texttt{Sambacry} are the main malicious cryptocurrency miner families 
that were actively targeting Windows and Linux OS from the year 2009 onward.


\texttt{BitcoinMiner.J} and \texttt{BitCoinMiner.hxao} use social engineering 
and phishing techniques to infect machines, while \texttt{Mal/Miner-c} and 
\texttt{Adylkuzz} use a worm component to spread and exploit other machines 
in the same network. Specifically, \texttt{Mal/Miner-c} exploits a design flaw 
in Seagate Central device~\cite{sophos} to infect other machines in the same network, and
\texttt{Adylkuzz} uses the exploit for Microsoft SMB vulnerability~\cite{eternalblue} 
dumped by the Shadow Brokers in the beginning of 2017. Similarly, \texttt{SambaCry} targets 
Linux machines exploiting a vulnerability in an older version of Samba~\cite{samba}.

\subsubsection{Coin mining}
~~The \texttt{Mal/Miner-c} malware contains three components that can be used for mining Monero (XMR) coins: 
\emph{NSCpuCNMine32.exe}, \emph{NSCpuCNMine64.exe} and \emph{NSGpuCNMine.exe}. After inspecting the 
CPU type and GPU of the victim's device, it selects the most suitable one to mine coins efficiently. 

Through manual analysis and code comparison, we discovered that, in general, the miner 
components are just obfuscated versions of CPUminer and are freely 
available\footnote{https://Bitcointalk.org/index.php?topic=647251.0}.
Prior to mining, the miner downloads a  configuration file from the C\&C server.
This file contains pool server details that the miner should use 
and the address of the wallet to which the mined coins should be credited. 
Moreover, the miner uses the Stratum protocol to communicate with the pool 
server of a mining pool. These mining pools are legitimate mining pools that can be used by anyone 
willing to contribute their computation power to mine coins. 

The cryptocurrency miner version of \texttt{Adylkuzz} surfaced 
only around May 2017. Through manual analysis of all available 
samples, we discovered that the miner component of this malware is a packed version of
CPUminer (version 2.3.3), which is open source. After the infection 
has taken place, the miner contacts the C\&C server for the configuration 
file, which contains information on what coin to mine and which pool server to use. 
This miner was also mining (privacy-friendly) Monero coins by connecting to a legitimate Monero 
pool server.  Moreover, it had two versions of the miner embedded in the binary, one for 32-bit systems
and another for 64-bit systems.

The \texttt{BitcoinMiner.hxao} cryptocurrency miner uses social engineering and phishing 
techniques to infect users. After infecting the device, the main component of the malware 
installs the miner component as a Windows background service. 
By reverse engineering the available samples, we discovered that this miner component is 
\emph{MinerGate Admin edition}\footnote{https://minergate.com/downloads/admin}, which is mining software for MinerGate mining pool users. Even though 
the pool supports various altcoins, the malware is configured to mine the same untraceable Monero coins.
Anyone can register to this pool using a valid email address. Then, to 
mine a cryptocurrency, the user can run this mining software on any device 
by just specifying her email and the pool URL. Finally, the user can withdraw 
the mined coins from her wallet address by signing into her pool account. While we found 
the email address that is being used by these malware samples, without knowing the
password of the account, we cannot find how much Monero this group of cyber criminals managed to mine.

The \texttt{Sambacry} \cc{} miner takes over Linux servers to also mine Monero coins by exploiting 
a Samba remote code execution 
vulnerability. After successful execution on the server, it installs a backdoor which 
gives the attacker shell access to the server.  Using this access, the attacker uploads 
and execute two files: \emph{INAebsGB.so} and \emph{cblRWuoCc.so}. The INAebsGB.so 
file makes sure that the attacker has a 
persistent access to the machine.  Meanwhile, the other file - cblRWuoCc.so, downloads the open source 
version of CPUminer. The wallet address where the mined coins should be added 
and the pool server that should to be used by the miner were
hard-coded in the binary. Using these information, an analyst could find how many coins this malware mined so far.

\subsubsection{Cash out}
~~Prior research~\cite{sophos} on \texttt{Mal/Miner-c} aimed 
 to find out how many coins the malicious miner was able to mine 
using the APIs provided by the Monero pool server.
According to the author, the mining pool paid 58,577 Monero (XMR) to the attacker's wallet. 
Note that, when the malicious attack was launched, the price of one XMR 
was 1.5 USD and at the time of writing this paper (September 1, 2017), one 
XMR was valued at 140 USD---making the value of the mined XMR
by this campaign some 8,200,780 USD, today.

Moreover, an independent investigation on \texttt{BitcoinMiner.hxao}~\cite{minerhxao} 
showed that a campaign managed to mine 2,289 XMR which is equivalent to 320,460 USD today.   

Another study~\cite{proofpoint} analyzed \texttt{Adylkuzz} and estimated that a campaign which 
took place in 2017 managed to mine 1,570 XMR using 3 different wallet addresses in a period of 
three weeks. However, there is a high probability that there might be more campaigns that are 
using different wallet addresses. Hence there is no certainty about the total coins mined by 
\texttt{Adylkuzz}. Today, the value of the Monero coins mined by this campaign is at least 
220,000 USD.

Lastly, according to~\cite{sambacry}, \texttt{SambaCry} mined 98 XMR in a month using a 
single wallet address. To date, the value of the Monero coins mined by this campaign is 13,720 USD.

\subsection{Cryptocurrency mining on Android}
\texttt{BadLepricon} and \texttt{Kagecoin} were two cryptocurrency miner families targeting Android and whose apps were found in the Google Play store disguised as wallpaper apps. Additionally, 
a version of \texttt{Kagecoin} spread through third-party stores as a repackaged app.

\subsubsection{Coin mining}

~~Based on the samples from our dataset, we found that \texttt{BadLepricon} was 
mining Bitcoins whereas \texttt{Kagecoin} focused on altcoins 
such as Litecoin, Dogecoin and Casinocoin. Different cryptocurrencies use 
different proof-of-work (PoW) algorithms, hence the apps need to implement 
these algorithms in order 
to mine for that particular coin. All the mining apps for Android available 
today are using the open source version of CPUminer that supports different PoW algorithms. 
Moreover, in the case of Android, we observed that to reduce the overhead, 
cyber criminals combine it with open source CPU miner code for ARM, which they 
embed in apps as native code. 

After fulfilling the advertised functionalities of the app, 
\texttt{BadLepricon} enters into an infinite loop where every five seconds it checks the battery level, network connectivity, and the display status (to see if the phone's display is on). While it almost seems as if it performs these checks as a 
courtesy to the user's phone, in reality it helps the malware to fly under the radar and survive longer. Firstly, when left unsupervised, mining algorithms can damage a phone by using excessive processing power which ultimately burns out the device. In order to avoid this, \texttt{BadLepricon} makes sure that the battery level is running 
at over 50 percent capacity, the display is turned off, 
and the phone network connectivity is on. Secondly, monitoring the phone's battery status is a 
good way of hiding your activities. \texttt{BadLepricon} 
also uses the WakeLock permission---an Android feature that makes sure the 
phone does not go to sleep even when the display is turned off.

For \texttt{Kagecoin}, the miner component starts as a background 
service. Since cryptocurrency mining is a CPU intensive operation, it could run down the battery of the device very quickly. The first 
version of \texttt{Kagecoin} (ANDROIDOS\_KAGECOIN.HBT) did not have 
any mechanism to hide this behaviour from the user. However, the second version 
of the malware only mines when the phone is in its charging 
state, so that most users will be oblivious to any suspicious activities taking place on their device.

\subsubsection{Cash out}

~~In order to control millions of bots, the malware author 
may use a proxy to set up one point of contact. \texttt{BadLepricon} 
uses a Stratum mining proxy~\cite{stratumproxy} that allows an attacker 
to change mining pools dynamically. This makes it difficult for 
 analysts to find out the attacker's wallet address.
 Even so, \texttt{Kagecoin} uses a different strategy---it has a 
component to update the configuration of the bot which 
describes the coin to be mined, the attacker wallet address 
and the pool to which to connect. In either case, the attacker consolidates the mined coins in a 
single wallet and then exchanges them for money using cryptocurrency exchanges.

\subsection{Cryptocurrency mining in the IoT}

While going through the samples downloaded from VirusTotal, we found a family of malware explicitly targeting Raspberry Pi 
devices. This malware was first uploaded to VirusTotal in August 2017.  Kaspersky classified it as \texttt{Shell.Miner}. 
As a side note, in 2014 Symantec reported a malicious cryptocurrency miner version called \emph{Darlloz}\footnote{https://www.symantec.com/connect/blogs/iot-worm-used-mine-cryptocurrency} which also targetted 
IoT devices---more specifically, IP cameras. However, we were unable to find a sample of this malware for our analysis today.


To facilitate its propagation, \texttt{Shell.Miner} scans the IoT network for Raspberry Pi devices  
that are using default usernames and passwords.  After pwning the device, it 
installs the open source version of CPU miner and then changes the password.


\subsubsection{Coin mining}

~~Through manual analysis, we found that \texttt{Shell.Miner} 
uses a popular open source miner called \emph{cpuminer-multi} to mine Monero \cc{}. 
We also found the pool information and the attacker's wallet address which are hardcoded in the 
shell script.  The pool server used by this malware is \url{xmr.crypto-pool.fr} (Figure~\ref{fig:shellminer}).



We used a feature provided by the pool server to find out how many XMR were rewarded to this 
address by the mining pool. Figure~\ref{fig:shellminer} shows that the botnet is still 
actively mining Monero coins for the attacker at the time of writing this paper.  We 
contacted the administrator of this pool and asked them to block the suspicious wallet 
address. It should be noted that this botnet is generating 2400 H/s (hashes per second), where one Raspberry 
Pi can only generate 8 to 10 H/s. This means that for this wallet, the botnet comprises some 
3000 machines or more that are also being targeted by the same cyber criminal group.

\subsubsection{Cash out}

\begin{figure}
    \center
\includegraphics[width=1\linewidth]{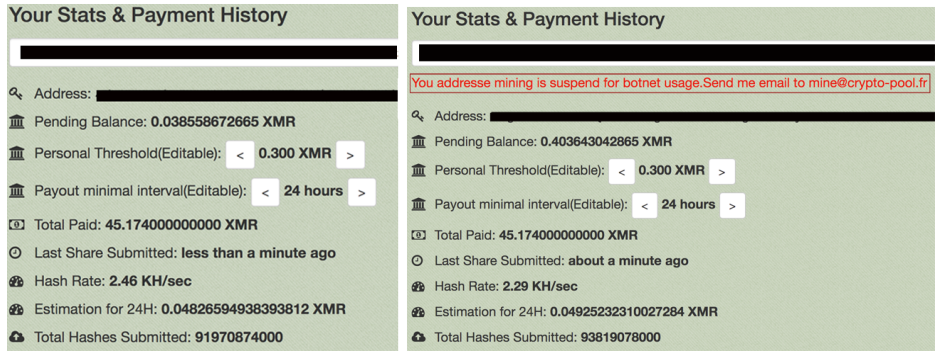}
\caption{Although the  pool admin blocked payouts to the reported wallet address, the botnet still mines Monero---not for the cyber criminals, but for the pool.}
    \label{fig:shellminer}
  \end{figure}

~~Figure~\ref{fig:shellminer} shows that the criminals managed to mine 45 XMR with an 
accumulated hash rate of 2400 H/s. Since Monero's dollar value as of 
01 September 2017 was 140 USD, this means the attacker received 6,300 USD 
in three months and paid to this wallet address.  

\setlength{\textfloatsep}{0.4cm}

For contrast, Symantec reported that \texttt{Darlloz} mined 
42,438 Dogecoins (approx. 46 USD in 2014) and 282 Mincoins 
(approx. 150 USD in 2014)  using an open source miner software. 
It should be noted that the value of these cryptocurrencies and 
mining difficulty of Dogecoin was very low in 2014, and that 
the price of Dogecoin increased by 976 percentage within three years. 
Therefore, it is only logical and most rewarding for the cyber criminals 
to mine coins with very low difficulty and keep it in the wallet till the currency value goes up.

\section{Outlook}
\label{sec:outlook}


\paragraph{\textbf{The advantages of malicious mining}} ~~Malicious cryptocurrency mining started as early as
2009---the year that Bitcoin was introduced. In the years that
followed, most criminals focused on banking trojans such as
SpyEye and (GameOver) Zeus. But mining malware also gained
popularity among cyber criminals who were, after all, already in the
business of compromising PCs and herding large numbers of them in
botnets.


Meanwhile, other criminal activities also rose in popularity,
especially \emph{ransomware}---malware that encrypts a user's data and
holds it hostage until the user pays the ransom money. Compared to
banking trojans, they were less visible to the banks and the
police. Specifically, the ransom amounts were small and all money
transfers were legitimate as far as the banks were concerned, and
initiated by the users themselves. From previous
studies~\cite{ransompayout}, we know that around 3 percent of the
victims pay the ransom---some 300 USD in Bitcoin. Thus, the scheme
revolves around the infection of many machines storing valuable user
data. Moreover, the payout per infection is a one-time amount of a few
hundred dollars. After that, many bots are `burned' as users typically
clean up their systems to prevent further damage. Additionally, ransomware 
is noisy: if many people get infected, many
people talk about it~\cite{locky,wannacry,nonpetya}. Attracting
attention is not always good for the criminals.

No such limitations exist for cryptocurrency mining: a bot can
easily mine less common privacy-preserving cryptocurrencies to make all
transactions untraceable from the start. Modern
miners are now starting to target exactly these kinds of cryptocurrencies.

In addition, unlike the one-time hit delivered by traditional
ransomware or banking trojans, cryptocurrency miners are designed to
fly under the radar and generate a \emph{continuous} stream of revenue
for their botmasters. They also do not need to infect machines that
contain valuable data or that are used for online banking---every
infected machine immediately contributes to generating
profits. Moreover, the miners steal money indirectly, as victims never
transfer money directly to a criminal's account. Instead, they pay via
their energy bills and the performance loss of their devices. As a
result, the damage is much less visible.

Finally, a growing wave of new cryptocurrencies make it profitable to
mine coins without specialized hardware, even if this is no longer
feasible for currencies such as Bitcoin.  
While
cryptocurrencies appear and disappear continuously, the cryptocurrency
market capitalization is increasing
exponentially~\cite{ElBahrawyAKPB17}, and altcoins make up a sizable
portion of the market (over 50\% at the time of writing) and are often
still CPU mineable. In fact, we will show that even low-end IoT
devices are interesting targets for \cc{} mining.  

Given the advantages and the increase in mining malware families in
the wild, a further proliferation of malicious \cc{} mining may
perhaps be expected.




\paragraph{\textbf{Altcoin mining}} 
~~Our study of malicious miners
in the previous section shows that cyber criminals try to mine the most lucrative coin (easy to mine, untraceable and valuable) at that specific point in time.  For instance,  \texttt{BitcoinMiner.hxao},  \texttt{Mal/Miner-c}, \texttt{Adylkuzz} and \texttt{Sambacry} and  \texttt{Shell.Miner} are mining Monero precisely because it is CPU mineable, untraceable and valuable. Table~\ref{table:cpualtcoins} shows the list of CPU mineable coins that are actively traded today.
Of course, doing so, requires keeping an eye on the specific altcoins to mine, since both the mining difficulty and the exchange rates change over time.


\begin{table}[t]
  \caption{CPU Mineable altcoins}
  \label{table:cpualtcoins}
  \centering
\begin{scriptsize}
  \begin{tabular}{|l|l|l|}

    \hline

\textbf{Coin} 		& \textbf{Algorithm}	& \textbf{Exchanges} \\
    \hline
Boolberry	&	Wild Keccack	& Poloniex, Bittrex	\\
JackpotCoin	&	JHA	& Bittrex, MintPal, Poloniex	\\ 
MemoryCoin	&	Momentum with SHA-512	& Bter, Poloniex	\\ 
YACoin	&	Scrypt-YAC	& Cryptsy	\\
Moneta Verde	& CryptoNight & Poloniex \\
New Universal Dollar	&	Bcrypt & 	Bittrex	 \\
Riecoin	&	Prime Constellations	 & Poloniex, MintPal, BTC38 \\
Slimcoin	&	Dcrypt	& Bter	\\ 
Darkcoin	&	X11	& MintPal, Poloniex, Bter \\
Givecoin	&	X11	& Bittrex	 \\ 
Global Denomination	&	X11	& Poloniex, Bittrex, AllCrypt \\
Hirocoin	&	X11 &	MintPal, Poloniex, Bittrex	\\ 
Logicoin	&	X11	& Poloniex \\
GroestlCoin	&	Groestl & 	Poloniex, MintPal \\
Aeon	&	CryptoNight	& Poloniex, Swaphole	\\ 
Bytecoin	&	CryptoNight &	Poloniex, HitBTC \\
Ducknote	&	CryptoNight &	Poloniex, Bittrex \\
Fantomcoin	&	CryptoNight &	Poloniex, Bittrex	\\
Monero	&	CryptoNight	& Poloniex, Bittrex, MintPal	\\ 
QuazarCoin	&	CryptoNight &	Poloniex, Bittrex, Swaphole	\\ 
    \hline

  \end{tabular}
\end{scriptsize}
\end{table}

\paragraph{\textbf{Infection vector}}
~~Similar to any other class of malware, the malicious miner can use any method to spread---phishing, drive-by-download, malvertising, social engineering or worm propagation. 
In the beginning of 2017, several families of ransomware used worms as an infection vector. For instance, the WannaCry ransomware infected more than 300 thousand devices, while Mirai and BrickerBot infected millions of IoT 
devices. Now malicious miner families have 
started copying  this method of spreading to reach as many  victims as possible.
Another approach (actively discussed on underground forums, these days)  is to buy bots for 
negligible prices on the underground markets and utilize them for 'malicious mining' purposes.
Finally, a recent report~\cite{malvertising} showed that browser-based mining malware now spreads via malvertising. In this case, the criminal buys the traffic from advertisement networks to serve JavaScript-based mining scripts to users' browsers (instead of serving ads).
This form of mining is known as \emph{web mining}.
Later in this section, we discuss the profitability of this business model.


\paragraph{\textbf{Devices targeted}} ~~Our study shows that currently most of the malicious miners are targeting high-end machines. Nevertheless,  we also see significant mining on low-end machines, and here, the criminals are targeting IoT devices more than mobile devices, even though mobile devices usually have more computation power.  The ease of infection and the high likelihood of staying hidden, are probably the main reasons.  With a worm component, it is easier to infect hundreds of thousands or even millions of IoT devices.  For instance,  Mirai and BrickerBot infected up to tens of millions of IoT devices~\cite{mirai,brickerbot}. We also saw \texttt{Shell.Miner} generating 2400 H/s. A recent report by Intel on the IoT forecasts 200 billion connected devices by 2020~\cite{inteliot}. Most of these IoT devices have stable internet connectivity, even  compared to mobile devices---enabling the attacker to mine on them throughout the day. Finally, for an average user, it is harder to detect a miner running on her security camera or Raspberry Pi device than to detect a malicious app on her mobile devices.

\paragraph{\textbf{Mining Components}}
~~Another interesting finding is that all the malicious miners are
using a free version of legitimate CPUminer software. The open source
version of CPUminers are optimized to give maximum results, and from a
profit perspective it thus makes senses that malicious miners also make use of 
these solutions. Moreover, all miners are using the
same pool communication protocol, i.e. the \emph{Stratum} protocol. In our
investigation, we did find that several families  tried to obfuscate the mining components, presumably to evade signature-based detection.  
\begin{table}[t]
  \caption{Miner component used by malware in our dataset}
  \label{minercomponent}
  \centering
  \begin{tabular}{|l|l|c|}
    \hline
    \textbf{Malware family} & \textbf{Miner component} & \textbf{Obfuscated version}\\
    \hline
    Mal/Miner-C & Claymore's CryptoNote CPU Miner v3.5 &  Yes\\
    BitcoinMiner.hxao  & MinerGate Admin edition &  No\\
    Kagecoin  & CPU Miner 2.3.2  & No \\
    Adylkuzz & CPU Miner 2.3.3 \footnote{https://Bitcointalk.org/index.php?topic=632724} &  Yes\\
    SambaCry & CPU Miner 2.3.3  & No  \\
    Shell.Miner  & CPU Miner Multi  & No \\
    \hline
  \end{tabular}
\end{table}

\paragraph{\textbf{Profit}} ~~When it comes to profit, the specific altcoin to mine plays a crucial role. At the time of writing, our analysis shows that Monero altcoin is the best candidate to mine on high-end machines, and on IoT devices with a stable internet connection. ByteCoin (BCN) is a good candidate to mine on mobile devices because it has comparatively lower difficulty than Monero, while still offering significant monetary value, as well as privacy.

A PC can yield somewhere between 30 H/s and 300 H/s if it is mining a CryptoNote-based cryptocurrency. Table~\ref{benchmark} shows the result of our experiments on different devices. A botnet usually consists of PCs with different computation power. Let us conservatively assume 
a botnet consist only of low-end PCs that can generate at least 30 H/s each. In that case, each  bot generates 
at least .0007 Monero coins per day~\cite{cryptocompare,minergate}.  With 10,000 bots, this botnet generates at least 854 USD per day. Similarly, if a botnet in the  IoT consists of devices that generate 9H/s, each device contributed some 7.5 BCN per day.  
Similarly,  for IoT devices let's assume, on average, one device generates 9 H/s.  This means a device can generate at least 7.5 BCN. On 01 September 2017, the value of BCN was 0.0024 USD which means it is possible to make at least 180 USD per day using a botnet of 10 thousand devices with similar computational power. Note that such botnet sizes are considered very small.

Let us now consider the case where cyber criminals hire a botnet for malicious cryptocurrency mining which at least in theory, is an easy to set-up criminal business model. While these botnets are generally offered as----ironically named---booters and stressors to perform DDoS-attacks, some botnets also offer the possibility to use them for other purposes. Two separate studies by respectively Verisign and WebRoot show that it is possible to rent botnets and they discuss pricing on specific underground markets~\cite{webroot,verisign}. In the study done by Verisign, analyzing 25 underground market botnet renting offerings, they conclude that renting out a botnet of 10 thousand bots costs 67 USD per 24-hour window.  This means, renting 10 thousand bots for 67 USD to mine Monero offers a very good profit margin, assuming the botnet use is exclusive.


Finally, let us estimate the profit that can be gained by \emph{web mining}. Coinhive~\cite{coinhive} is a legitimate online service which offers a JavaScript miner for the Monero blockchain that webmasters can embed in their website to monetize the visitors' computation power.  At the time of writing, Coinhive provides 0.00016414  XMR per 1M hashes, while charging 30\% fees. We can use this service as a basis for evaluating this business model. From experiments, we find that a PC-based browser yields 10 to 60 H/s depending on its computation power. So let us assume, on average, a PC yields 20 H/s which means 10 thousand users visiting a website for 5 minutes generates 60M hashes which are equivalent to  1.3 USD. In comparison, a website can generate 2.7 times more money by displaying one popup ad to 10 thousand visitors\footnote{http://paypopup.com/advertising\_packages.php}. It would seem that web mining is only profitable for websites that bind many visitors for a long time, such as perhaps sites that stream movies or sports events, or that host online games, as discussed in~\cite{minesweeper}. 

\paragraph{\textbf{Cashing out without getting caught}} ~~Some malicious cryptocurrency miners   use a configuration file hardcoded in the binary that specifies things like  the cryptocurrency to mine,  the pool server to use, and the wallet address to credit. Others use a dynamic configuration file which they download from a control and command server at start-up time. In the latter case, the botmaster has a  chance to update the configuration file, e.g.,  with a new wallet address in case the old wallet address gets blocked by the mining pool. Interestingly, some mining pools provide APIs to find out how much cryptocurrency it has  paid to a particular wallet address. For instance, we used these APIs to find out how many coins  \texttt{Shell.Miner} had mined. After discussing with two pool administrators from two different pool servers, we realized that pool admins may not care about cyber criminals using botnets for cryptocurrency mining, as in any case, it increases the pool's profit. Interestingly, even though one pool administrator eventually blocked payouts to the wallet address that we found in the \texttt{Shell.Miner} binary, the botnet is still mining Monero coins---not for the cyber criminals, but for the pool (Figure~\ref{fig:shellminer})!

In order to not get blocked by the mining pool, some malicious miners, such as \texttt{BadLepricon}, use proxy pool server~\cite{stratumproxy}. In this case, bots contact the proxy server to get jobs and submit shares. The proxy server forwards the requests and shares to the appropriate pool server with the attackers'€™ credentials (including the wallet address). Thus, an analyst studying the malware cannot find the attackers' wallet address at all, as this is only specified at the proxy server, under the control of the attacker.

Another approach to avoid  getting flagged by authorities is to use a mining pool that does not provide any details about its users. For instance, we showed that \texttt{BitcoinMiner.hxao} uses the MinerGate pool by specifying only the email address of the attacker on the victim'€™s device. Analysts cannot find how many coins this botnet mined without knowing the password of that account, or without the support from the pool administrators. The analyst can only report the email address to the administrators, hoping they will at least block the payout to that account. 
Moreover, our research shows that even with the support from pool administrators, it is impossible to catch the cyber criminals if they are mining untraceable coins like Monero.

If for some reason (e.g., because they hope that  the value of a specific coin will explode in the future), cyber criminals insist on mining a cryptocurrency that provides limited privacy, they can still trade the mined currencies for privacy-preserving currencies using anonymous exchanges such as \emph{Shapeshift}~\cite{shapeshift}. Such exchanges allow anyone to trade cryptocurrency instantaneously without registering with them.  Depending on the market value of the coin, the attacker can then cash out from any exchange without any legal problems. 





\begin{table}[t]

  \caption{Device benchmark : Coin value fluctuates}
  \label{benchmark}
  \centering
\resizebox{\linewidth}{!}{%
  \begin{tabular}{|l|l|c|c|}
    \hline
    \textbf{Device type} & \textbf{CPU} & \textbf{Hashes per second} & \textbf{Bytecoin per day} \\ 
    \hline
       & AMD FX-8350 Eight core 4.0GHz & 320 & 250 \\ 
PC & Intel Core i7-3770k 3.5 Ghz & 250 & 195 \\ 
       & Intel Core i5-3470 3.2GHz & 170 & 134 \\ 
\hline
       & Quad-core (2x2.15 GHz Kryo \& 2x1.6 GHz Kryo) & 20 & 15 \\ 
Mobile & Quad-core 2.3 GHz Krait 400 & 15 & 11 \\ 
       & 1.2 GHz quad-core Cortex-A7 & 8 & 6 \\ 
\hline
IoT &   Raspberry Pi 3 & 9 & 7 \\ 

        \hline
  \end{tabular}}
\end{table}

\paragraph*{\textbf{Detection}}
\label{sec:detection}
~~Detecting most of the current \cc{} miners is fairly straightforward, and our study identified various conspicuous features, such as 1) the reuse of open source version of miners, 2) the clear-text communication protocol,  and 3) specific  permissions and native code usage in Android miners. All of these could be used to flag a system as suspicious. However, we can not rely on these characteristics  for the future, as malware authors can easily change their code to not exhibit them. Instead, we should look for more fundamental properties that would be extremely hard for malware authors to hide.  In particular, properties such the recurring high CPU load,  power consumption,  and heat generation. Also, for some devices (e.g., on the IoT), we could detect suspicious communication patterns. Measuring these properties may help an anomaly detector determine that a  \cc{} miner is present.

\section{Related Work}
\label{sec:relatedwork}

Previous studies~\cite{botcoin:ndss2014,mcafee_threat_report:2014} on cryptocurrency mining claim that mining is less profitable than other malicious activities, such as spamming or booter-renting (DDoS for hire) and should be used as a secondary monetizing scheme. These studies focus on botnets that were used to mine Bitcoin during the year 2011--2013.  There are no reports of these botnets being used for cryptocurrency mining in the last couple of years, which implies they may not be active anymore.  However, our research focuses on all types of malicious cryptocurrency miners, including web miners that are targeting three different platforms -- PCs, mobile devices and IoT devices. We also show how this threat evolved along with the evolution of the cryptocurrency market and reached a state where malicious cryptocurrency mining has become a highly profitable, and relatively safe and silent strategy for cybercriminals. 

A recent report~\cite{minerhxao} from Kaspersky supports our finding by showing that the number of cryptocurrency miner attacks detected in the first eight months of 2017 seems on track to exceed that of 2016. Another study~\cite{malvertising} shows that a new cryptocurrency mining malware is found to be using malvertising as the infection vector. However, we show that \emph{web mining} through malvertising is not profitable in the case where victims spend very little time online. 

Finally, a different yet related line of work, MineGuard~\cite{mineguard}, shows that it is possible to detect covert mining operations in cloud platforms. MineGuard serves as a hypervisor tool based on hardware-assisted behavioral monitoring, which accurately detects the signature of a miner. Similar signature-based detection can be used to detect malicious cryptocurrency miners targeting PCs.

\section{Conclusion}
\label{sec:conclusion}

In this paper, we examined the phenomenon of malicious cryptocurrency
mining. Compared to just a few years ago, such activities have become
more interesting to cyber criminals. We analyzed the current families
of these malicious miners on several platforms, as well as the
developments in the cryptocurrency markets that have removed some of
the barriers for malware to profit from mining. We have seen that even
compromised low-end devices can now lucratively mine cryptocurrencies
for criminals. The result of our study shows that it makes
economic sense (again) for attackers to invest in this type of activity.
We cautiously predict that we may expect a proliferation of
such attacks in the near future.

\section*{Acknowledgements}

This research was supported by the European Union's Marie Sklodowska-Curie grant agreement 690972 (PROTASIS).  
This project further received funding from  MALPAY consortium, consisting of the Dutch national police, ING, 
ABN AMRO, Rabobank, Fox-IT, and TNO. Any dissemination of results must indicate that it reflects only the authors'
view and that the Agency is not responsible for any use that may be made of the information it contains.

\bibliographystyle{splncs03}
\bibliography{bibliography}

\end{document}